# Infra-red reflectance and emissivity spectra of nanodiamonds

A Maturilli[1], A A Shiryaev[2,*], I I Kulakova[3], J Helbert[1]

[1] Institute for Planetary Research, DLR, Berlin, Germany

[2] Institute of Physical chemistry and electrochemistry RAS, Leninsky pr. 31, 119071 Moscow, Russia

[3] Chemistry department, Moscow State University, Russia

* - corresponding author: shiryaev@phyche.ac.ru OR a_shiryaev@mail.ru

**Abstract**

Reflectance and emissivity spectra of nanodiamonds powder were measured in a dedicated setup at temperatures up to 873 K. The spectra are characterised by presence of sharp bands due to surface-bound functional groups. Thermal desorption of oxygen-containing groups lead to corresponding spectral changes. The maximal emissivity of nanodiamond powder reaches 0.985.

## 1. Introduction

Nanodiamonds obtained by detonation of explosives with a negative oxygen balance in a closed volume possess properties making them interesting for broad range of fundamental aspects and diverse applications (Dolmatov, 2007; Ultrananocrystalline Diamond, 2012). A remarkable feature of the detonation nanodiamonds is the relatively narrow size distribution with sharp peak around 5 nanometers. Despite intense work many of their spectroscopic properties are still not fully explained. One of the most interesting problems is the investigation of size-dependent changes of absorption and emission spectra. Whereas total emissivity of nanoparticles and clusters is reasonably well predicted theoretically (e.g. Martynenko and Ognev, 2005, Smirnov, 1997), experimental results for well-documented nanoparticulate systems are scarce. Currently, examination of the mere applicability of the Planck's law to nanomaterials is being actively investigated (Fan et al., 2009; Wuttke and Rauschenbeute, 2012).

Besides interest for solid state spectroscopy, knowledge of emissivity spectra of nanoparticles are important for astrophysical applications, since IR spectroscopy is used to infer the existence of nanodiamonds in various astrophysical environments. Usually the conclusions about the presence of nanodiamonds are based on observations of two prominent absorption bands due to C-H bond vibrations which are observable on surfaces of macrodiamonds or powders with grain sizes exceeding 50 nm (Guillois et al., 1999). However, similar bands are also observed in many hydrocarbons and thus can not serve as a solid proof that nanodiamonds are observed. In addition, nanodiamonds found in meteorites are considerably smaller, having median size around 2.6 nm (Lewis et al., 1987, Shiryaev et al., 2011). Therefore, measurements of emission spectra of small diamond particles can be important for a correct assignment of bands in astrophysical observations.

Though literature on spectroscopic properties of diamond is immense, experimental data about its emissivity is scarce. Emission spectra of macroscopic diamonds in relatively narrow wavelength range relevant for diamond applications as optical windows for $CO_2$ lasers were reported by Mollart and Lewis (2001). It was shown that the emissivity ($\varepsilon$) spectra closely follow the expected behaviour 1-R, where R is the reflectivity. Reflectivity spectra of diamond are presented in Phillipp and Taft (1962); emissivity and reflectance of macroscopic graphitic carbon are discussed for example, in Ergun (1968). In the current paper the first measurements of infra-red (IR) emissivity spectra of nanodiamonds are presented.

## 2. Experimental

Commercially available detonation nanodiamonds (UDA-SF trade mark, Joint Stock Co. "Almaznyi Tsentr," St. Petersburg, Russia) subjected to standard chemical treatment using 50-60% nitric acid at 230-240 °C and 80-90 atm were studied. Reflectivity and emissivity spectra of the nanodiamonds have been acquired at the Planetary Emissivity Laboratory (PEL) of DLR in Berlin using a dedicated setup described in details in (Maturilli at al., 2011). The measurements are performed with a Bruker Vertex 80V Fourier Transform Infrared spectrometer, coupled with en external emissivity chamber. Both systems can be evacuated to pressures lower than 1 mbar to remove atmospheric features from the spectra. An induction heating system heats the uniform 3 mm thick smooth layer of nanodiamonds samples poured in stainless steel cup to temperatures of up to 700 K, while the surroundings remain cold. The emissivity chamber is equipped with high precision temperature sensors for the samples and other internal parts, plus a webcam for taking pictures or videos of the sample cups under investigation. To cover the 1-16 μm spectral range, a liquid nitrogen cooled HgCdTe (MCT) detector and a KBr beamsplitter are used. An identical cup is filled with the calibration target, which in case of high temperature measurements is a blast furnace slag, since no blackbody chemical paint coating can survive temperatures higher than 570 K for prolonged periods. Each cup is equipped with a thermocouple with metal Inconel mantel possessing accurate readings to at least to 1270 K. The temperature readings are used as input to the induction controller, which automatically tunes the energy given to the pan-cake coil to reach the requested sample temperature. Once the sample cups are prepared on the rotating carousel, the emissivity chamber is closed and slowly evacuated to prevent dusting. Upon reaching stable chamber pressure, the induction is switched on, allowing approximately 1 hour waiting time to reach a constant desired temperature. The sample and the blackbody reference target are measured at identical temperatures by rotation of the carousel. This allows precise calibration of the emissivity of the samples, once corrected for the effective emissivity spectrum of the reference blackbody.

Spectra of nanodiamonds were measured at room temperature on fresh sample and after slow step heating to 773 K, when early stages of graphitisation have occurred, which was manifested by some darkening of nanodiamonds. The reflectivity of the samples was measured in air, since our experience shows that the measurements in air and vacuum give identical results provided correction for contribution of atmospheric species such as $CO_2$. For comparative purposes reflectance spectrum of graphite particles with sizes 10-32 microns was also recorded.

## 3. Results and discussion

Reflectance and emissivity spectra are shown in figure 1. A broad band centered at approx. 3200 cm$^{-1}$ and relatively sharp features around 3675, 2960, 2920, 2850, 1800, 1760, 1625 cm$^{-1}$ are observed in the reflectance spectra (fig. 1a) of all the samples correspond to functional and OH-groups chemisorbed on the surfaces of diamond and graphite particles. The fresh nanodiamond sample is characterised by slightly lower reflectivity that the graphite and graphitised nanodiamonds. Small changes between the samples are also clearly visible; the most important differences consist of a decreased amount of OH-groups after heating and variations in the configuration of >CH$_2$ groups and of the sp$^2$-coordinated C-H bonds. The maximum value for reflectance of the fresh nanodiamond sample is approx. 0.985.

The reflectance spectra of nanodiamonds show clear temperature evolution. One immediately observes gradual disappearance of sharp bands in the range 1000-1900 cm$^{-1}$ with heating (fig. 1b). In IR absorption spectra of nanodiamonds this spectral region is dominated by various functional groups present on grain surfaces (e.g., Jiang and Xu, 1995; Lisichkin et al., 2006). According to thermal desorption studies (Koshcheev et al. 2008, Koshcheev 2008) the most common carboxyl groups are destroyed in the range between 200-400 °C, lactonic and anhydride groups are somewhat more stable (desorption between 400-700 °C). Therefore, the changes in emissivity spectra are logically explained by progressive desorption of the surface groups during vacuum heating. Persistence of H-related features is related to their high binding energy to diamond surface (Koshcheev et al. 2008, Koshcheev 2008, Hamza et al., 1990, Shiryaev et al., 2007).

The emissivity spectrum acquired at the highest temperature employed (600 °C) differs from the others and apparent emissivity exceeds unity. In addition, some spectral features change sign. This irregularity is due to peculiar hydrodynamic behaviour of heated nanodiamonds layer as a medium with high Reynolds number. During the heating the initially smooth nanodiamonds layer has deteriorated and numerous holes appeared. Through these holes hot metal sample holder became visible. Its strong thermal emission dominated the spectrum, making comparison with the reference black body not very reliable (Hapke 1993, Salisbury 1993). The peculiar hydrodynamic behaviour of nanodiamonds layer is currently being investigated in detail.

This work shows that reflectance and emissivity spectra of nanodiamonds are similar, but not identical. This can be explained by inevitable porosity in the nanodiamond layer, dramatically complicating the interaction of radiation with the sample (Hapke, 1993; Salisbury, 1993). Clearly, the surface scattering regime dominates, whereas the volume scattering/absorption are negligible due to very small grain size in comparison with the wavelength. Sharp spectral features observed in both types of spectra are dominated by the contributions from surface functional groups. The emissivity spectra acquired in this work show that the existence of small heated nanodiamonds may indeed be inferred from IR astrophysical observations of spectral features related to C-H vibrations. In the same time, such assignment should be performed with certain caution due to the general similarity of this spectral region for various carbonaceous substances.

It is of certain interest to consider the use of nanodiamonds as a reference black body for emissivity spectroscopy. Nanodiamonds are relatively cheap and accessible nowadays and are stable to approx. 750 K. As shown in this work their reflectance is low (0.985), being attractive for application as a reference. However, practical experience shows that nanopowder layers are prone to dusting during chamber evacuation, thus limiting the speed of pumping. Another problem encountered in this work is an unusual behaviour of the heated layer of nanodiamonds, leading to formation of holes and other types of heterogeneities during heating. These irregularities are clearly detrimental for the wide use of nanodiamonds as a reference black body. However, sintered nanodiamonds might be considered for such applications.

**Acknowledgments**: This work is partially supported by RFBR grant 12-05-00208 (to AAS).

**References**

Dolmatov V Yu 2007 Detonation-synthesis nanodiamonds: synthesis, structure, properties and applications. *Russ. Chem. Rev.* 76, 339–360.

Ergun S 1968 Optical studies of carbon. In: Chemistry and physics of carbon, vol. 3, ed. P.L.Walker Jr., Marcel Dekker, New York. 45-120.

Fan Y, Singer S B, Bergstrom R, and Regan B C 2009 *Phys.Rev.Lett.*, 102, 187402

Guillois O, Ledoux G, and Reynaudet C. 1999 *Astrophysical J.* 521, L133

Hamza A V, Kubiak G D, Stulen R H 1990 *Surf. Sci.* 237 35

Hapke B 1993 Theory of Reflectance and Emittance Spectroscopy. Cambridge University Press, Cambridge.

Jiang T and Xu K 1995 Carbon 33 1663

Koshcheev A P 2008 *J. Russ. Chem. Soc.*, 52(5) 88

Koshcheev A P, Gorokhov P V, Gromov M D, Perov A A, Ott U 2008 *Russ. J. Phys. Chem.* A82(10) 1708

Lewis R S, Anders E and Draine B T 1989 Properties, detectability and origin of interstellar diamonds in meteorites. Nature 339 117.

Lisichkin G V, Korol'kov V V, Tarasevich B N, Kulakova I I, and Karpukhin A V 2006 *Russ. Chem. Bull.*, 55(12) 2212.

Martynenko Yu V and Ognev L I 2005 *Techn. Phys.* 50(11) 1522.

Maturilli A, Helbert J and D'amore M 2011 *42nd Lunar Planet. Sci. Conf.,* Abstract #1693 (http://www.lpi.usra.edu/meetings/lpsc2011/pdf/1693.pdf)

Mollart T P and Lewis K L 2001 *Phys.Sstat. Sol.* A186(2) 309

Phillipp H R and Taft E A 1962 *Phys.Rev.* 127 159

Salisbury J 1993 Mid-infrared spectroscopy: laboratory data. In: *Remote Geochemical Analysis*, ed. C. Pieters and P. Englert. Cambridge University Press

Shiryaev A A, Grambole D, Rivera A, Herrmann F 2007 *Diam. Relat. Mater.* 16 1479

Shiryaev A A, Fisenko A V, Vlasov I I, Semjonova L FP. Nagel P, Schuppler S 2011 *Geochim. Cosmochim. Acta* 75 3155

Smirnov B M 1997 *Phys. Uspekhi* 40 1117

Ultrananocrystalline Diamond 2012 2nd Edition, Eds.: Shenderova O A and Gruen D M, Elsevier

Wuttke C and Rauschenbeute A 2012 arXiv:1209.0536v1.

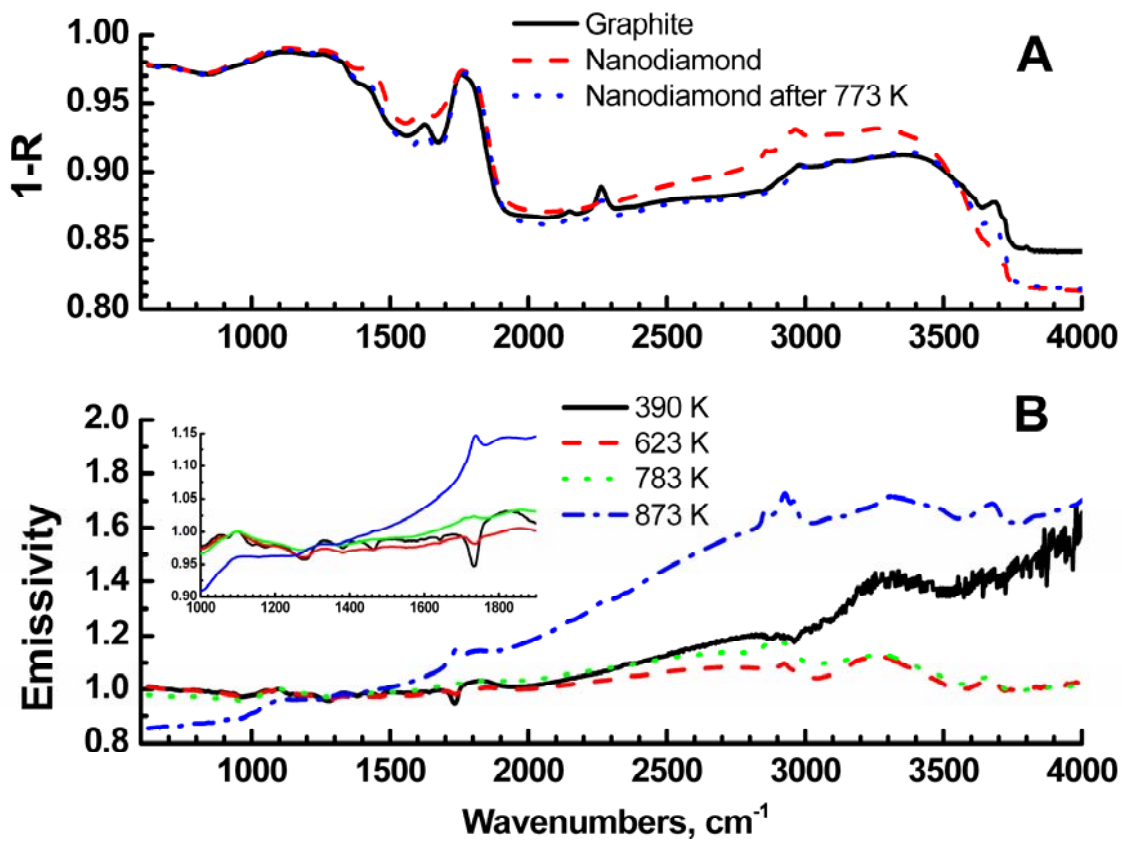

Figure 1. Reflectance (A) and emissivity (B) spectra of nanodiamonds at various temperatures and after heat treatment. Reflectance spectrum of microcrystalline (10-32 microns) graphite is shown for comparison. Inset in B shows zoomed part of the reflectance spectra of the range dominated by functional groups. See text for explanation of emissivity exceeding unity.